\documentclass[conference]{IEEEtran}
\IEEEoverridecommandlockouts
\usepackage[hidelinks]{hyperref} 
\hypersetup{ 
    colorlinks=false, 
    pdftitle={Consumer Group Autoscaler}
}
\usepackage{cite}
\usepackage{caption}
\usepackage{amsmath,amssymb,amsfonts}
\usepackage{float}
\usepackage[caption=false, font=footnotesize]{subfig}
\usepackage{algorithm, algorithmic}
\usepackage{mathtools}
\usepackage{xpatch}
\usepackage{xcolor}
\usepackage{listings}
\usepackage{realboxes}
\usepackage{verbatim}
\definecolor{codegreen}{rgb}{0,0.6,0}
\definecolor{codegray}{rgb}{0.5,0.5,0.5}
\definecolor{codepurple}{rgb}{0.58,0,0.82}
\definecolor{backcolour}{rgb}{0.95,0.95,0.92}
\lstdefinestyle{mystyle}{
    backgroundcolor=\color{backcolour},   
    commentstyle=\color{codegreen},
    keywordstyle=\color{magenta},
    numberstyle=\tiny\color{codegray},
    stringstyle=\color{codepurple},
    basicstyle=\ttfamily\footnotesize,
    breakatwhitespace=false,         
    breaklines=true,                 
    captionpos=b,                    
    keepspaces=true,
    numbers=left,                    
    numbersep=5pt,                  
    showspaces=false,                
    showstringspaces=false,
    showtabs=false,                  
    tabsize=2
}
\makeatletter
\xpretocmd\lstinline{\bgroup\appto\lst@DeInit{\egroup}}{}{}
\makeatother
\usepackage{graphicx}
\usepackage{textcomp}
\usepackage{xcolor}
\def\BibTeX{{\rm B\kern-.05em{\sc i\kern-.025em b}\kern-.08em
    T\kern-.1667em\lower.7ex\hbox{E}\kern-.125emX}}
\begin{document}

\title{
    Kafka Consumer Group Autoscaler
}

\author{
    \IEEEauthorblockN{Diogo Landau}
    \IEEEauthorblockA{
        \textit{Department of Electrical Engineering} \\
        \textit{University of Porto - FEUP}\\
        diogo.hewitt.landau@hotmail.com
    }

    \and
    \IEEEauthorblockN{Xavier Andrade}
    \IEEEauthorblockA{
        \textit{Department of Industrial Engineering} \\
        \textit{University of Porto - FEUP}\\
        xavier.andrade@fe.up.pt
    }
       \and
    \IEEEauthorblockN{Jorge G. Barbosa}
    \IEEEauthorblockA{
        \textit{Department of Informatics Engineering} \\
        \textit{University of Porto - FEUP, LIACC}\\
        jbarbosa@fe.up.pt
    }
}

\maketitle

\begin{abstract}
Message brokers enable asynchronous communication between data producers and consumers in distributed environments by assigning messages to ordered queues. Message broker systems often provide with mechanisms to parallelize tasks between consumers to increase the rate at which data is consumed. The consumption rate must exceed the production rate or queues would grow indefinitely. Still, consumers are costly and their number should be minimized. We model the problem of determining the required number of consumers, and the partition-consumer assignments, as a variable item size bin packing variant. Data cannot be read when a queue is being migrated to another consumer. Hence, we propose the R-score metric to account for these rebalancing costs. Then, we introduce an assortment of R-score based algorithms, and compare their performance to established heuristics for the Bin Packing Problem for this application. We instantiate our method within an existing system, demonstrating its effectiveness. Our approach guarantees adequate consumption rates – something the previous system was unable to – at lower operational costs.
\end{abstract}


\section{Introduction}

Message Brokers are a common tool employed to fixture communication in a
distributed environment \cite{magnoni2015modern}. This system  provides with asynchronous communication
between producers and consumers, and handles some of the challenges that are
common within distributed and concurrent data processing. 

We consider a queue or partition to be a structure within a message broker environment where messages are appended in the same order as they were produced. It is also important that this structure is capable of delivering the messages in the same order as they were produced. This consumption model is often a requirement to guarantee state consistency between two distinct distributed systems for a specific business related entity. Common applications that employ replaying a set of messages in the same order as they were produced include event-carried state transfer, and system state simulation. The former consists of having two systems reading the same set of messages in the same order to replicate an entities state within their data store \cite{burckhardt2012eventually, su2005slingshot}, whereas the latter is used to simulate a system's state at a given point in time by reproducing the messages up until that point \cite{carbone2017state, kshemkalyani1995introduction}.

\begin{figure}[htb!] 
    \centering
    \includegraphics[width=0.45\textwidth]{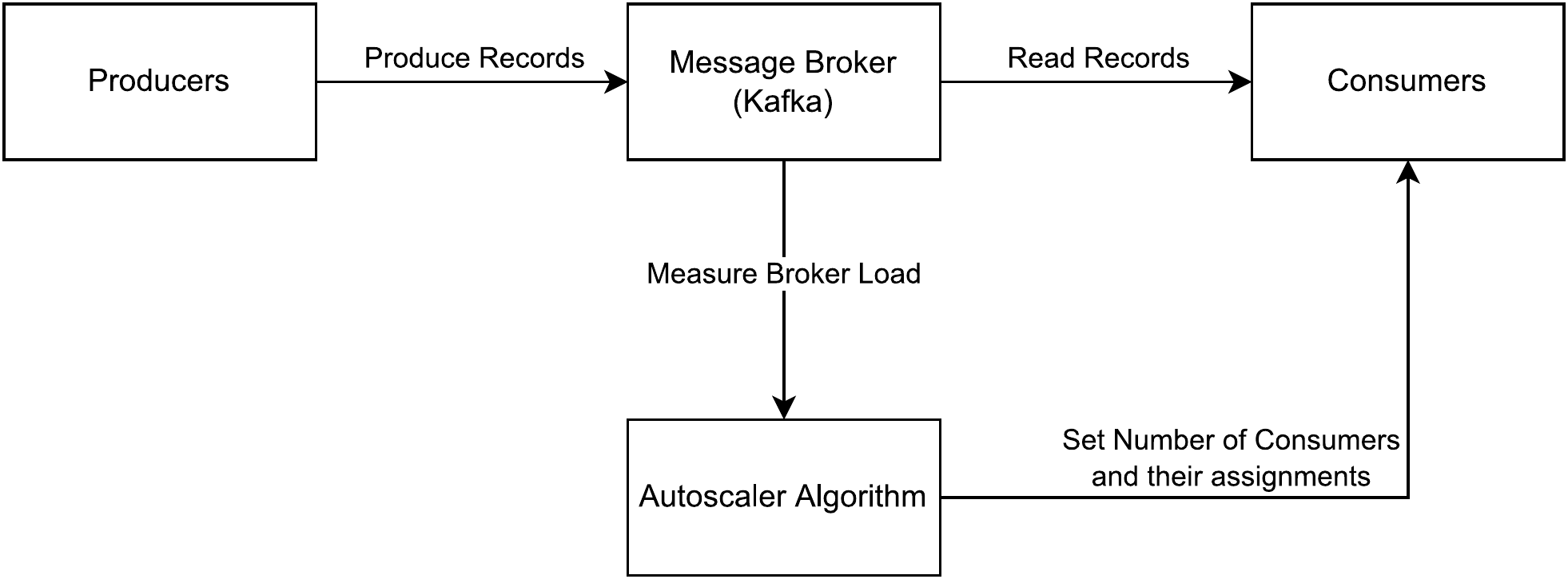}
    \caption{
        On request, our algorithm determines the number of consumers and partition-consumer assignment based on the message broker's current load. 
    } 
    \label{fig:high_level_representation} 
\end{figure}

Two common message broker implementations, Kafka and RabbitMQ, have different ways of guaranteeing message order on consumption. Within Kafka, all the messages produced to a partition are appended and delivered in the same order as they were published. Messages can be consistently produced to the same partition if the same key is used\footnote{Kafka Message Ordering: \url{https://kafka.apache.org/documentation/\#intro_concepts_and_terms}}. As for RabbitMQ, it guarantees messages toward a queue are enqueued in the same order as they were produced. Therefore, to guarantee same order message delivery, only a single consumer can be reading messages from the queue\footnote{RabbitMQ Message Ordering: \url{https://www.rabbitmq.com/queues.html\#message-ordering}}.

Within a distributed message broker environment, provided we are given a set of independent queues, our goal is to manage a group of consumers, so as to guarantee the rate of consumption is at least as high as the production rate. 

Since message ordering is important, only one consumer can be assigned a queue, but multiple queues can be assigned the same consumer. This also implies that the maximum number of consumers is limited to the number of queues.

As each consumer represents an active computing resource, while attempting to determine the amount of parallelism required, we are also aiming to minimize the operational cost, by reducing the number of consumers. 

Considering consumers as bins, and queues as items that have to be assigned a bin, this problem was modeled as a variation of the Bin Packing (BP) Problem, with the particularity that items vary in size over time. This occurs because a queue's size correlates to its current write speed, which fluctuates based on the current system's load and inevitably implies that a BP solution for a given time instant may not hold true in future instants. Figure \ref{fig:high_level_representation} provides with a high level representation of the interacting entities.

On account of this BP variation, a new solution has to be computed at different time instants, each having different information as to the size of each item (each queue's "current" write speed for that time instant). This might lead to a queue (item) being migrated to a different consumer (bin) when compared to the consumer group's previous configuration. 

Since two consumers cannot read from the same queue concurrently, there is another cost to take into account associated to rebalancing a partition. This cost is related to the amount of data that is not being read while the queue is being rebalanced.

Provided this is the first time the variable item size bin packing variant is applied in this context, existing algorithms do not take this rebalance cost into account. Hence, in Section \ref{sub:rscore} we propose a metric to account for a given iteration's rebalance cost (Rscore). Additionally, using the Rscore, in Section \ref{sub:modified_any_fit} we propose four new BP heuristic algorithms that account for the rebalance costs, three of which are shown in Section \ref{sub:exp_modified_any_fit} to be a competitive alternative to the multi-objective optimisation problem at hand.

Furthermore, it is our belief that this model is applicable in other contexts, e.g., the Kubernetes node autoscaling problem. In Kubernetes, nodes are machines that have resources, to which pods are assigned. Each pod is independent and the node resources it consumes varies with time. Therefore, modeling nodes as bins and pods as items whose size varies over time, this problem also fits our variation of the BP problem, with the added complexity of bins being allowed to have different sizes, which is not considered in our scenario. This problem presents similar characteristics to the work elaborated in \cite{wolke2015more} and \cite{song2013adaptive}.

In Section \ref{sec:related_work}, we present applications of the Bin Packing Problem, and a comparison of existing approximation algorithms to solve the Bin Packing problem. Section \ref{sec:problem_formulation}, formally defines the problem, followed by the proposed approach in Section \ref{sec:algorithm}. We also present in Section \ref{sec:autoscaler} an application within a message broker environment, more specifically Kafka, of autoscaling a group of consumers based on the broker's current load. We use Section \ref{sec:experimentation} to present and discuss the tests performed on the proposed algorithms, and the autoscaling model when applied to Kafka. Lastly, we conclude this paper through Section \ref{sec:conclusions}.

\section{Related Work}
\label{sec:related_work}

\subsection{Bin Packing Applications}

There are several applications where BP is used to provide with an optimal
or sub-optimal solution \cite{nardelli2019efficient, FURINI2012251, dell2020branch, delorme2017logic}. The Virtual Machine Placement (VMP) problem, which has
gained more attention due to increasing use of cloud providers to support
companies' technological infrastructure.
The problem can be generalized to a set of Virtual Machines (VM),
each having to be assigned one of the cloud provider's Physical Machines (PM) while
attempting to minimize the operational cost. A variation of BP arrises in this
scenario where the items are considered ephemeral and add another temporal
dimension since they have associated a start time and a duration. 

Cauwer et al. \cite{de2016temporal} study BP with a temporal variant, where the
tasks have an equal start time. Their aim is to minimize the unused capacity of
a physical machine over time, and therefore the optimal solution for the BP
problem does not necessarily provide the optimal solution in their Temporal
variant of the BP problem. In addition to the time model presented by
\cite{de2016temporal}, Ayd{\i}n et al. \cite{aydin2020multi} study the case
where items may have differing start times. Their objective function is more in
line with the traditional BP, as it aims to minimize the number of physical
machines (bins) used. Moreover, due to the increased energy costs related to machine
fire-ups (switching a machine off when it is not in use and back on again when
required), the authors also attempt to minimize the number of fire-ups, making
their problem multi-objective.

Furini et al. in \cite{furini2018matheuristics}, study a stateful BP variant,
where the time dimension is discretized into several time instants, each with tasks that have to be executing at that time, and therefore have to be assigned a PM. Their problem has an additional constraint, related to the
tasks having to run uninterrupted, and therefore cannot be assigned two
different bins for their execution period. Consequently, solving each discrete
problem independently does not provide with an optimal arrangement, as that
would require the items to be reassigned.

The variable item size variant of the bin packing problem, has also been studied by Wolke et al. \cite{wolke2015more} and Song et al. \cite{song2013adaptive}, due to the varying load patterns presented by VMs. Both periodically execute a reallocating controller that determines new PM assignments for each VM, while aiming to minimize the cost of having multiple PMs running, and the migration cost. The migration cost in their problem is related to a VMs memory footprint. 

Xia et al. \cite{xia2013throughput} considers the online bin packing problem within the context of Fog computing, to determine whether a cloudlet has the resources to offload a user's request.

Bin packing has also been applied to real-time data streaming, which is a new paradigm of data processing, where compute nodes process the data in real-time, as opposed to batch processing the data periodically. Further elaborating, when processing data in real-time, a query is executed "continuously" to provide with real-time insights as soon as a new tuple is provided. As Gulisano et al. \cite{gulisano2012streamcloud} point out, a continuous query can be construcuted as a direct acyclic graph, where the different operators can be distributed to different nodes, having upstream nodes communicating with their downstream conterparts to communicate the result of its operator. This also enables for a distributed execution of a query, and an elastic infrastructure capable of scaling based on the current system's load. The authors use utilization thresholds to determine how to scale its cluster of nodes. 

Heinze et al. \cite{heinze2013elastic}, use bin packing to dynamically scale in or out a set of nodes based on the current system's query load, while maintaining process latency constraints. The authors observed underutilization, due to query operators no longer occupying computing resources in host machines after task completion, and therefore, devised a re-balance policy. They acknowledge re-executing the bin packing heuristic would provide with better utilization rates, but in their scenario this was not feasible due to excessive re-balance cost, and therefore devise a re-balancing algorithm that re-places only the operators that present minimal load in their system, until a specific condition is met. As a complement to the previous findings, \cite{heinze2014auto} extends the algorithm to use reinforcement learning to improve the decision to scale out or in, based on the utilization rates and latency measurements. 

\subsection{Approximation Algorithms}
\label{sub:approximation_algorithms}

The Bin Packing (BP) problem is a well established research problem, and has been
extensively reviewed in the literature. Due to the time constraints imposed by
our applicatoin, this paper gives emphasis to the Approximation Algorithms
that heuristically solve the problem in low-order polynomial time, as opposed to the higher time complexity Linear Programming approach.

A method is presented to classify the BP problem in \cite{coffman2013bin},
which will be used throughout this section. During an algorithm's execution, a
bin can find itself either \textit{open} or \textit{closed}. In the former, the
bin can still be used to add additional items, whereas in the latter, it is no
longer available and has already been used.  

The list of bins is indexed from left to right, and the number of bins used by
an algorithm can be computed using the index of the first empty bin in the list of bins. When creating a bin, this process can be visualized as opening the lowest index of the empty bins (left-most empty bin).

Using $A(L)$ to denote the amount of bins a certain algorithm makes use of for a
configuration of items $L$, $OPT(L)$ to represent the amount of bins
required to achieve the optimal solution, and defining $\Omega$ as the set of
all possible lists, each with a different arrangement of their items,  
\begin{equation}
    R_A (k) = \sup_{L \in \Omega} \left \{ \frac{A(L)}{k} : k = OPT(L) \right \},
\end{equation}
encodes a performance metric relative to the optimal number of bins used by an
algorithm. The Asymptotic Performance Ratio (APR) of an algorithm $A$
($R_A^\infty$) is defined as
\begin{equation}
    R_A^\infty = \limsup_{k \to \infty} R_A(k).
\end{equation}
This is the metric that will be used throughout this section to compare the
algorithms' performance.

Despite the mulititude of classes presented in \cite{coffman2013bin}, the only
class of problems that is of interest for this paper, is the offline bin
packing variant.

Offline BP algorithms are aware of the list of items that are to be
assigned bins prior to its execution. As long as the set of items remains the
same, items can be grouped, sorted or anything that might be convenient to the
algorithm that is going to iterate over the list of items. 

The Any Fit subclass of algorithms must fit the constraint: \textit{if bin $j$ is empty, an item will not be assigned to it if the item fits into any bin to the left of $j$}. These algorithms perform best if the list of items is sorted in a non-increasing order prior to assigning bins to the items. The offline
approximation algorithms that are going to be described, follow this sorting
strategy, and all but the Next Fit Decreasing (NFD) belong to this subclass of
algorithms. Provided a list of sorted items, the NFD only has a single open bin
at a time, which is also the last created bin. This algorithm attempts to place
the item in the right-most non-empty bin. If it does not fit, then the bin to
the right is used. As shown in \cite{baker1981tight}, this algorithm has as APR,
\begin{equation}
    R_{NFD}^\infty \approx 1.6910.
\end{equation}

With regards to the Worst Fit Decreasing (WFD), for each item, it attempts to
assign the bin with most slack. If the size constraint is not satisfied, a new
bin is assigned.

The Best Fit Decreasing (BFD) is similar to the WFD differing only on its packing
strategy, wherein this algorithm attempts to place the item in the bin where it
fits the tightest. In case there is no open bin to fit the current item, a new
bin is created where the item is inserted. 

Lastly, the First Fit Decreasing (FFD) places each item in the left-most bin as
long as the bin's capacity is not exceeded. As shown in
\cite{johnson1974worst},
\begin{equation}
    R_{BFD}^\infty = R_{FFD}^\infty = \frac{11}{9}.
\end{equation}

In combination with their online variants, these algorithms will be used as a
reference to evaluate the performance of our Modified Any Fit algorithms in Sec.
\ref{sec:experimentation}.

\subsection{Conclusion}

Although the temporal aspect of the Bin Packing Problem has already been
reviewed, there is a gap when it comes to the possibility of rebalancing the
items between the different bin instances. Existing solutions lack a deterministic metric to account for the rebalancing cost, and present conservative heuristic algorithms to migrate the items to different bins (items with higher cost are not migrated). 

Also, in view of the fact that the existing approximation algorithms were developed to solve a single iteration of the BP problem, these have to be adapted to the problem at hand, which requires solving a new iteration given an already existing assignment of items to bins. Despite the fact that the adaptation presented in Section \ref{sub:approximation_algorithms_adaptation} improves these algorithms' performance with respect to the rebalance cost, there is still room for improvement. 

Therefore, since modeling the message broker consumption domain as a variable item size bin packing variant, is, to the best of our knowledge, a novel approach, in Section \ref{sub:rscore} we provide with a metric (Rscore) that encodes the rebalance cost for a given iteration. Based on this metric, we contribute with four new heuristic algorithms that simultaneously aim to minimize the operational and migration costs (Section \ref{sub:modified_any_fit}).

\section{Problem Formulation}
\label{sec:problem_formulation}

\begin{figure}[htb!] 
    \centering
    \includegraphics[width=0.45\textwidth]{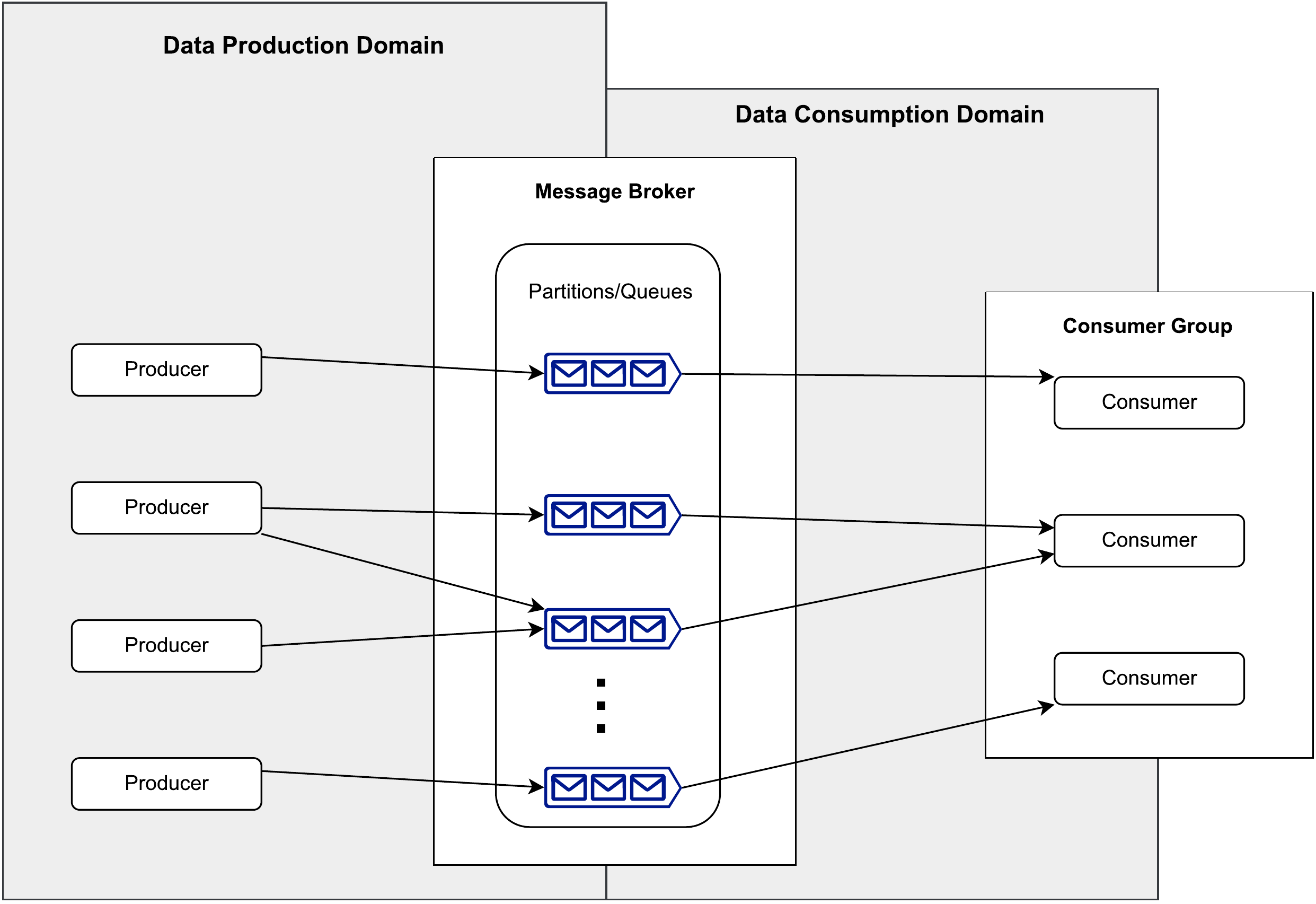} 
    \caption{
        Representation of data production and consumption domains within a message broker environment. 
    } 
    \label{fig:kafka_representation} 
\end{figure}

A group of consumers is interested in consuming data from several partitions/queues with varying load within the message broker. The set of partitions is distributed between the different consumers of the group to simultaneously parallelize and load-balance. Each partition can only be assigned a single consumer, whereas a consumer can be assigned several partitions. This problem is contained within the Data Consumption Domain of Figure \ref{fig:kafka_representation}. 

The aim of our work is to achieve a deterministic approach to determine the
number of consumers required working in parallel, so as to guarantee that the
rate of production into a partition is not higher than its rate of consumption
by the consumer group. In contrast, if this were not the case, messages would
accumulate leading to the lag between the last message inserted and the last
message read by the consumer group to increase with time.

We assume that the maximum consumption rate of a single consumer is constant,
and if there is enough data to be consumed, this is the speed the consumer
functions at when working "full throttle". This is elaborated in Section \ref{sec:experimentation} based on several tests performed to determine this capacity value for a consumer. 

Based on the previous information, the model for this pipeline fits the
constraints of the Single Bin Size Bin Packing (SBSBP) variant, where the bins
are consumers that have as capacity their maximum consumption rate, and the
weights are the partitions and their respective write speeds. The problem then
is to find the minimum amount of consumers where to fit all the partitions so as
to make sure that the sum of the write speeds of the partitions assigned to a
single consumer does not exceed its maximum capacity. 

For a single iteration, the optimal solution finds the arrangement between the
partitions and the consumers that minimizes the amount of instances, but this
solution is not static, as the write speeds of the partitions changes with time.
As such, there is a second factor to take into consideration which is the
partition reassignment. 

When a partition is reassigned, because two consumers from the same group cannot
be consuming from the same partition at a time, when assigning a partition to
another consumer, the one it is currently assigned to has to stop consuming in
order to allow the new consumer to start. Due to this process, there is some
downtime where data is not being consumed from a partition.

For this reason, making use of an optimal algorithm that also minimizes
partition redistribution, is not feasible as it would not run within the
necessary time requirements due to the NP-hard nature of the problem \cite{garey1978strong}. To determine the arrangement of partitions and consumers, this work is
based on approximation algorithms. As is clear, there is no approximation
algorithm which considers item redistribution. 

Consequently, the BP problem is remodeled to take the time instant $t$ at which
the algorithm is to be executed into account.
\begin{alignat}{3}
\label{BPP model}
    &\min       
        &&\sum_{i \in B} y_i 
            && \\
    &\text{subject to} \quad
        && \sum_{j \in L} \hat w_j(t) \cdot x_{ij} \leq y_i \quad      
            && \forall \; i \in B \label{opt:c1}\\
    &   && \sum_{i \in B} x_{ij} = 1, \quad                             
            && \forall \; j \in L \label{opt:c2}\\
    &   && y_i \in \{0, 1\}                                             
            && \forall \; i \in B \\
    &   && x_{ij} \in \{0,1\}                                           
            && \forall \; i \in B, j \in L, 
\end{alignat}
where set $B$ corresponds to the available bins, and $L$ as the list of items to
be arranged into different bins. Decision variable $x_{ij}$ and $y_i$, indicate
whether an item $j$ is packed in the bin $i$, and whether bin $i$ is used
respectively. As for Constraints \ref{opt:c1} and \ref{opt:c2}, the former
assures that the sum of the items in a bin does not exceed it's capacity, and
the latter makes sure that every item is assigned a bin. 

Time instant $t$ represents an instant at which a measurement was performed of
the items' sizes, leading to a new computation of the bin packing assignment,
and new values for $x_{ij}$ and $y_i$. 

Since an assignment represents a consumer reading data from a partition, an item
can also be reassigned to another consumer (bin), if at time instant $t$ there
is already a consumer from the group reading data from it.
\section{Algorithm}
\label{sec:algorithm}

\subsection{Rscore}
\label{sub:rscore}

Within this problem's context, rebalancing inevitably implies having the
consumer group stop consuming from a partition while it is being reassigned from
one consumer to another. There cannot be a concurrent read from the same
partition by members of the same group.

This paper introduces the Rscore metric to compute the total rebalance cost of a
group's reconfiguration. The metric reflects the impact of stopping the data consumption
from the partitions that are being rebalanced. As such, the Rscore computes the
rate at which data accumulates, in consumer iterations per second.

In essence, the combination of the time it took to rebalance a set of partitions
and the Rscore, encodes information as to the maximum number of consumer
iterations required to fully recover the data that accumulated, while the set of
partitions was being rebalanced.

\begin{table}[htb!] 
\centering 
\caption{Data required to compute the Rscore.} 
\label{tab:rscore_data}
    \begin{tabular}{ |p{0.05\textwidth}|p{0.4\textwidth}| } 
        \hline 
        \textbf{Notation} & \textbf{Description} \\ 
        \hline 
        $P_i$ & Set of partitions rebalanced in iteration $i$ \\ 
        $s(p)$ & Write speed of partition $p$ \\ 
        $C$ & Maximum consumer capacity in
            bytes/s\\ 
        \hline 
    \end{tabular} 
\end{table}

Provided the data presented in Table \ref{tab:rscore_data}, the following
Equation \ref{eq:rscore} presents $R_i$, the Rscore for iteration $i$.

\begin{equation} \label{eq:rscore}
    R_i = \frac{1}{C}\sum_{p \in P_i} s(p) 
\end{equation}

\subsection{Modified Any Fit}
\label{sub:modified_any_fit}

Any Fit algorithms focus on reducing the amount of bins used to pack
a set of items, disregarding the cost associated with reassigning items. These need to be modified to include the rebalancing costs that affect the solution performance significantly.
Algorithm \ref{alg:modified_any_fit} illustrates the implementation of the modified any fit that will be further described for the remainder of this section.


Given the current consumer group's state (the partitions assigned to each
consumer), and a set of unassigned partitions, the modified algorithms differ
from the decreasing versions of the Any fit algorithms described in Section
\ref{sub:approximation_algorithms}, wherein the former class of algorithms does not sort the set of items
but, in turn the consumers of a consumer group based on their assignment. 

There are two sorting strategies when sorting the group of consumers: 
\begin{itemize} 
    \item Sorting consumers based on the cumulative speed of all partitions
        assigned to it (cumulative sort); 
    \item Sorting consumers based on the partition assigned to it that has
        the biggest measured write speed (max partition sort).  
\end{itemize}

After sorting the current consumer group using one of the above strategies, for
each consumer in the sorted group, the partitions assigned are sorted based on
their write speed. From smallest to biggest, each partition is inserted into one
of the bins that have already been created for the next iteration's assignment (future assignment). The insertion is made based on
one of the any fit strategies, which can either be the Best or Worst Fit
strategy. If the insert is successful, the partition is removed from the
sorted list of partitions. Otherwise, if there is no existing bin that can hold
the partition, then the current consumer assigned to it is created. 

The remaining partitions in the sorted list are inserted into the
created bin (which is also the consumer they are currently assigned to), from
biggest to smallest. If the partition is inserted successfully, it is removed
from the list of partitions. If a partition does not fit into the newly created
consumer, then the remaining partitions are added to the set of unassigned
partitions.

After repeating this procedure for all consumers, there is now a set of
partitions which have not been assigned to any of the consumers in the future
assignment. The final stage involves first sorting the unassigned partitions in
decreasing order (based on their measured write speed), and assigning each partition to consumer from the new consumer group according to a fit strategy.

\begin{table}[htb!] 
\centering 
\caption{Modified implementations of the any fit algorithms.} 
\begin{tabular}{ |p{0.4\columnwidth}|p{0.5\columnwidth}| } 
    \hline 
    \textbf{Algorithm} & \textbf{Consumer Sorting Strategy} \\ 
    \hline
    Modified Worst Fit (MWF) & Cumulative write speed \\ 
    \hline
    Modified Best Fit (MBF) & Cumulative write speed \\ 
    \hline
    Modified Worst Fit Partition (MWFP) & Max partition write speed \\ 
    \hline
    Modified Best Fit Partition (MBFP) & Max partition write speed \\
    \hline
\end{tabular} 
\end{table}

\begin{algorithm}[H]
\caption{Modified Any Fit Algorithm}
\label{alg:modified_any_fit}
\begin{algorithmic}[1]
    \renewcommand{\algorithmicrequire}{\textbf{Input:}}
    \renewcommand{\algorithmicensure}{\textbf{Output:}}
\REQUIRE  
    current consumer group configuration $C$ \& \\ 
    currently unassigned partitions $U$
\ENSURE    
    new consumer group configuration $N$
\STATE {$N \gets $ new ConsumerList with assign strategy}
\STATE {$S \gets $ sort $C$ on cumulative or max partition}
\FOR {$c \in S$}
    \STATE $pset \gets $ partitions assigned to $c$
    \STATE $pset \gets $ sort $pset$ in decreasing order
    \FOR {$i \gets pset.size()-1$ to $0$}
        \STATE $p \gets pset[i]$
        \STATE $result \gets N.assignOpenBin(p)$
        \IF {$result = false$}
            \STATE $break$
        \ENDIF
        \STATE $pset.remove(p)$
    \ENDFOR
    \IF {$pset.size() = 0$}
        \STATE $continue$
    \ENDIF 
    \STATE $N.createConsumer(c)$
    \FOR {$p \in pset$}
        \STATE $result \gets N.assign(c, p)$
        \IF {$result = false$}
            \STATE $break$
        \ENDIF 
        \STATE $pset.remove(p)$
    \ENDFOR
    \STATE $U.extend(pset)$
\ENDFOR
\STATE $U \gets$ sort $U$ decreasing order
\FOR {$p \in U$}
    \STATE $N.assignBin(p)$
\ENDFOR
\RETURN $N$
\end{algorithmic}
\end{algorithm}

\subsection{Approximation Algorithms Adaptation}
\label{sub:approximation_algorithms_adaptation}


Adding to the procedures described in Sec. \ref{sec:related_work}, another step was included in the approximation algorithms when creating a new bin that does not have an affect on the total number of bins used by these algorithms. If the consumer that is currently assigned to the partition has not yet been created in the future assignment, this is the bin that is created, otherwise, the lowest index bin that does not yet exist is the one created. This is not presented as a modification to the existing algorithms, it simply adapts these to the added rebalance concern.

\section{Kafka Consumer Group Autoscaler}
\label{sec:autoscaler}

\begin{figure*}[htb!] \centering
    \includegraphics[width=0.9\textwidth]{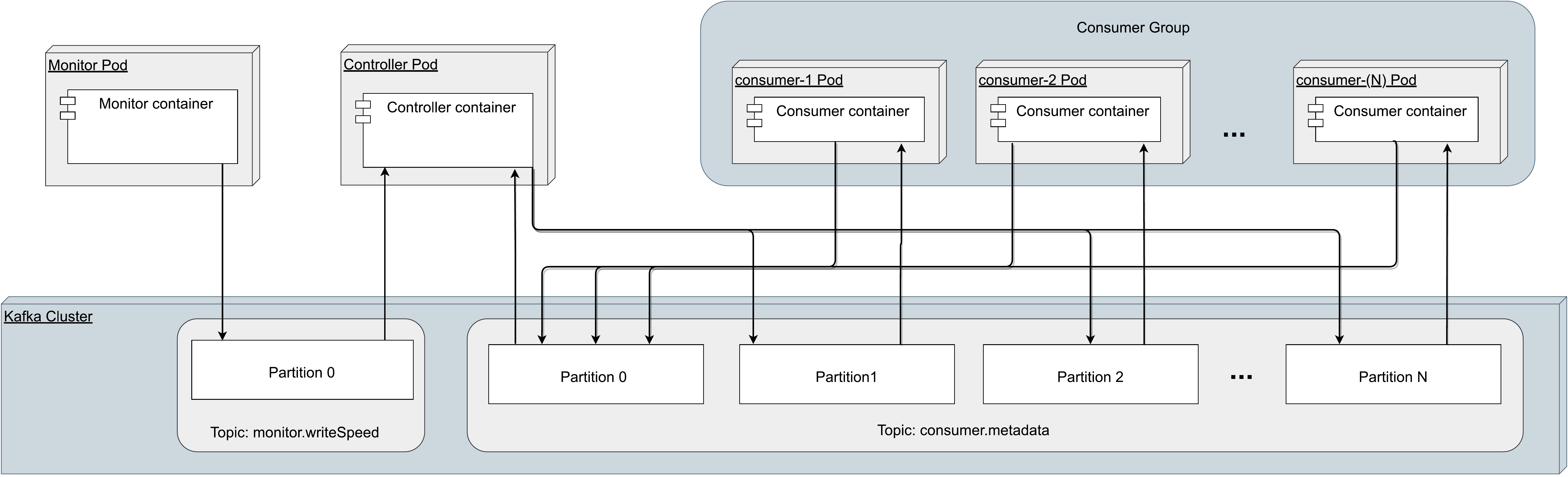}
    \caption{System architecture.} 
    \label{fig:system_architecture}
\end{figure*}

There are three components that interact with one another to provide a fully dynamic pipeline capable of autoscaling
based on the current load of data being produced to the partitions of interest,
as presented in Fig. \ref{fig:system_architecture}. 

\begin{enumerate}
    \item The monitor process is responsible for measuring the write speed of each
partition the group is interested in consuming data from, which is equivalent to
specifying the size of the items of the bin packing problem. 
    \item This information is then delivered
to the controller, which is responsible for managing a consumer group (creating
and deleting consumer instances), and mapping each partition to a consumer.
    \item The consumers are then informed of their tasks and read the data from the partitions
that were assigned to them by the controller.
\end{enumerate}

This system was applied in a real world context, with the goal of extracting data published to Kafka topics to be inserted into a Datalake, being a part of the Extraction domain of the ETL process. 

\subsection{Monitor}

To solve the BP problem, initially, the controller requires as input the write
speed of each partition the consumer group is interested in consuming data from.
The monitor process is responsible for providing this information.

Kafka provides an Admin client, which can be used to administer the cluster and
query information about it. This client/class exposes a method
$describeLogDirs()$ which queries the cluster for the
amount of bytes each TopicPartition has. A TopicPartition is a string-integer
pair, which identifies any partition (integer) within a topic (string). 

Each time the admin client queries the partition size, a timestamp is
appended to the measurement, and it is inserted into the back of a queue. Any query older than $30$ seconds, which is guaranteed to be in the front of the
queue, is removed. We use the average write speed over the last 30 seconds to estimate a single partition's write speed. We compute the difference between the first and last queue elements (i.e., the earliest and latest measurement of the partition size) regarding the last $30$ seconds. We divide the result by that time amount to obtain our write speed estimate in $bytes/s$.

After computing the write speed for all the partitions of interest, the
information has to be communicated to the controller/orchestrator. To benefit
from an asynchronous approach, this monitor process communicates with the
controller/orchestrator process via a Kafka topic illustrated in Fig. \ref{fig:system_architecture} as \lstinline{monitor.writeSpeed}. 

\subsection{Consumer}
\label{sub:consumer}

\begin{figure}[!htb] 
    \centering
    \includegraphics[width=0.40\textwidth]{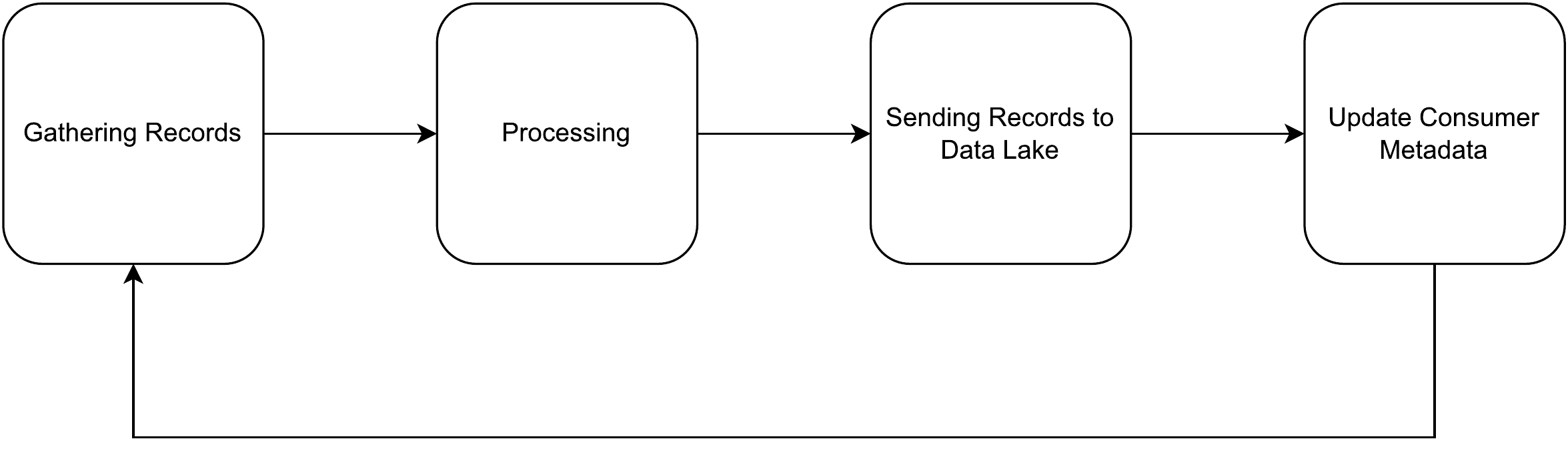}
    \caption{Consumer Insert Cycle.}
    \label{fig:consumer_cycle} 
\end{figure}

The Consumer goes through four important phases within its process to
approximate its consumption rate to a constant value when being challenged to
work at its peak performance. These phases repeat cyclically until the consumer
is terminated by an external signal. Fig. \ref{fig:consumer_cycle}
illustrates the consumer's process.

The consumer is configured with two important parameters,
$BATCH\_BYTES$ and $WAIT\_TIME\_SECS$, which indicate
respectively, the amount of bytes the consumer waits to gather in a single
iteration, and the amount of time it is allowed to wait to gather the
information. 

The first phase has the consumer attempt to fetch $BATCH\_BYTES$ from
the partitions it is consuming data from. After this condition is satisfied or
$WAIT\_TIME\_SECS$ is exceeded, the consumer moves on to the second
phase. Here the consumer processes each individual record, and batches records
that originated from the same topic together. There are different data lake
tables for each topic, and therefore not all records are inserted into the same
destination table.

The third phase is where the consumer sends the records into the data lake,
performing an asynchronous request for each topic it fetched data from in the
same iteration.

Having forwarded all the records into their respective data lake tables,
the consumer then verifies its metadata queue to verify if there are any change
in state messages to be consumed. This queue is how the Controller process (Sec.
\ref{autoscaler:controller}) informs each consumer of their assignments. If
there are new messages, the consumer reads all messages in the queue, and
updates its state. Only after having successfully updated its state and
persisted its metadata, does the consumer send an acknowledgment back to the
Controller to indicate the successful change in state. This cycle repeats
until the Controller removes the consumer from its group of active consumers. 

\subsection{Controller}
\label{autoscaler:controller}

The controller is the system component responsible for
orchestrating and managing the consumer group. The write rate of each partition computed
by the monitor process is an input to the approximation algorithm that determines new consumer group configurations. Based on these
configurations, the controller creates the consumers that do not yet exist,
communicates changes in assignment to each consumer in the group, and
deletes consumers that are not required in new computed group's states.

As shown in Fig. \ref{fig:system_architecture}, there are two topics that
function as communication intermediaries between the three system components.
The topic illustrated by \lstinline{monitor.writeSpeed} is where the controller
reads the partitions' write speeds, whereas the topic illustrated by
\lstinline{consumer.metadata} is where the controller sends messages to each
consumer to inform the change in their state. Regarding the latter Kafka topic,
partition 0 is reserved for communication directed toward the controller,
whereas the remaining partitions within \lstinline{consumer.metadata} represent
a one-to-one mapping to each consumer. Therefore, if the controller wants to
communicate with consumer $N$, it sends a record into metadata partition $N$.

This communication architecture was devised to achieve an efficient
communication model, defined as the amount of information that is relevant
relative to the amount of data read. With this pattern, all the data
read is relevant to the reading entity.

\begin{figure}[!htb] 
    \centering
    \includegraphics[width=0.45\textwidth]{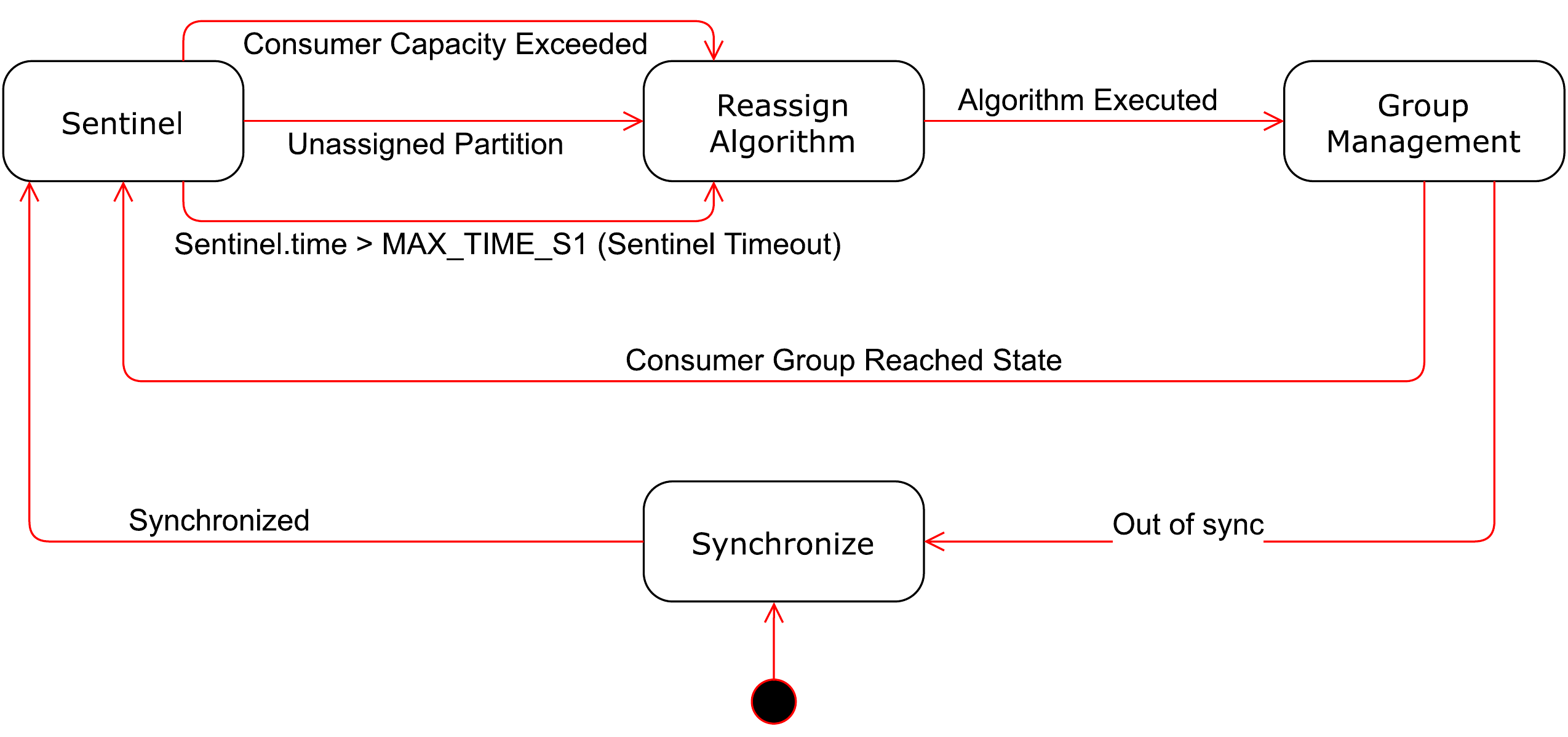}
    \caption{Controller State Machine.} 
\label{fig:state_machine} 
\end{figure}

The controller can be summarized by a state machine, intended to continuously
manage a group of consumers and their assignments, illustrated by Fig.
\ref{fig:state_machine}. 

The Sentinel state has the controller read the information published by the
monitor component to update its current consumer group's state, and determine
whether it has to recompute the group's assignments. If it evaluates one
of the exit conditions as true, the controller transitions into the
Reassign Algorithm state. Then, it employs one of the approximation algorithms
mentioned in Sections \ref{sub:approximation_algorithms} and \ref{sub:modified_any_fit} to solve the Bin Packing
Problem heuristically. 

The approximation algorithm outputs the desired state in which
the controller wants its group of consumers. Hence, the first step within the
Group Management state is to compute the difference between the current state of
the group and the desired state. This difference encodes information regarding:
the consumers to be created; the partitions each consumer has to stop consuming
from; the partition each consumer has to start consuming from; and lastly, the
consumers that have no assignment, and therefore can be decommissioned. 

The state difference allows the controller to operationalize the transition into the new state.
Firstly, the controller creates the new consumer instances, which involves the controller
communicating with a Kubernetes cluster to create the deployments related to
each consumer. We chose to have different deployments for each consumer, so that
the information related to which $consumer.metadata$ partition the
consumer had to read from, could be transmitted through the deployment's
manifest $metadata.name$ attribute. Secondly, the controller informs
each consumer of their change in state by transmitting the start and stop
consuming messages to each consumer via their respective
$consumer.metadata$ partition. It is important to note that
communicating the state change is a synchronous process. This assures that while
rebalancing partitions there at most one consumer from the group reading
data from a partition. This means that for each partition being rebalanced, the
controller first sends out the stop consuming message. Only after the
consumer informs the controller of having acted upon the message, can the
controller send out the start consuming record to the new consumer. After the
consumers have reached their intended state, the controller shuts down the
inactive consumers. 

The Synchronize state is used to synchronize the consumer group's real state
with the controller's perceived state. This is to guarantee the controller is
capable of recovering from an unexpected termination, which can lead to an
inconsistent perception of the consumer group's state.
\section{Experimentation}
\label{sec:experimentation}


\subsection{Test Data Generation}
\label{sec:random_data_generation}

Throughout this section, we use the term measurement to represent a map of
values that indicate the speed for each partition of interest. The term
stream refers to a sequence of measurements in the form of a list. Each
measurement of a stream simulates the items' sizes for an instance of the BP
problem, which is an input for the approximation algorithms. 

\begin{table}[H] 
\centering 
\caption{Data to generate a stream.} 
\label{table:testing_data} 
\begin{tabular}{ |c|l| } 
    \hline 
    \textbf{Symbol} & \textbf{Description} \\ 
    \hline 
    $P$ & Set of partitions of interest for the consumer group. \\ 
    $s_i(p)$ & Speed for a partition $p \in P$ at an iteration $i$ \\ 
    $\phi(\delta)$ & Uniform random function that selects a value between $[-\delta, \delta]$\\
    $C$ & Bin Capacity for the Bin Packing Problem.  \\
    \hline 
\end{tabular} 
\end{table}

To generate a stream of measurements, given $N$ (the number of measurements
desired) and $\delta$ (maximum relative speed variation between two sequential
iterations), at first the initial speed $s_0(p), \; \forall p \in P$ has to be
defined. Four different approaches were tested for the partition's initial
value: choosing a random value between $[0, 100]\% \cdot C$; setting the initial
speed of all partitions to $0$; setting the initial speed of all
partitions to $50\% \cdot C$; setting the initial speed of all partitions to
$C$. Given that there was no significant difference on the outcome
with these variations, the results presented were obtained using the streams of
data generated having an initial speed for all partitions that was randomly
selected between $[0, 100]\% \cdot C$.

Therefore, given $s_0(p)\ \forall p \in P$, the remaining measurements were
obtained using: 
\begin{equation}
\begin{split}
    s_i(p) = &  \max\{
                    0, 
                    s_{i-1}(p) + \frac{\phi(\delta)}{100} \cdot C
                \}, \\
             & \forall \ p \in P \ \wedge \ 
                         i \in \{1, 2, ..., N-1\}
\end{split}
\end{equation}

Using the aforementioned procedure, 6 different streams of data were generated
by setting $N = 500$ and setting $\delta$ to a value belonging to the set $\{0,
5, 10, 15, 20, 25\}$ for each stream of data ($\delta$ does not change within a
stream).

\subsection{Metrics}

The metrics used to compare the performance between the algorithms are the
Cardinal Bin Score ($CBS_\delta(a)$) and the Average Rscore ($E_\delta^a(R)$)
over all iterations of each stream.

\begin{table}[H] 
\centering 
\caption{
    Data to compute the cardinal bin score and the expected Rscore for a stream of measurements.
} 
\label{table:bin_score} 
    \begin{tabular}{ |c|p{0.8\columnwidth}| } 
    \hline 
    \textbf{Symbol} & \textbf{Description} \\ 
    \hline 
    $A$ & Set containing the existing and proposed approximation algorithms. \\ 
    \hline 
    $z_i^\delta(a)$ & 
        number of bins used in iteration $i \in \{0, 1, ..., N-1\}$ of a stream
        defined by $\delta$, by algorithm $a \in A$. \\ 
    \hline 
    $R_i^a$ & 
        Rscore for an iteration $i \in \{0, 1, ..., N-1\}$ for the outcome
        provided by algorithm $a \in A$. \\ 
    \hline 
\end{tabular} 
\end{table}

The cardinal bin score is calculated using the following expression:
\begin{equation}
\begin{split}
    CBS_\delta(a) = 
       &\frac{1}{N}
            \sum_{i=0}^{N-1} 
                \frac{  
                    z_i^\delta(a) - min_{b \in A} \{z_i^\delta(b)\} 
                }{
                    min_{b \in A} \{z_i^\delta(b)\} 
                }, \\
        &\forall \ a \in A \ \wedge \ \delta \in \{0, 5, 10, 15, 20, 25\}. 
\end{split}
\end{equation}
This metric expresses how many more bins, on average, an algorithm $a \in A$ has
compared to the algorithm that made use of the least bins, which encodes the
operational cost. The lower this value, the better the algorithm performed for a
given stream.

The expected value of the Rscore, is used to compare the rebalance cost for a
single stream, and is computed as follows:
\begin{equation}
\begin{split}
    E_\delta^a(R) = 
        &\frac{1}{N} 
            \sum_{i=0}^{N-1} R_i^a, \quad \\
        &\forall \ a \in A \ \wedge \  \delta \in \{0, 5, 10, 15, 20, 25\}.
\end{split}
\end{equation}
The aim is also to minimize an algorithm's Average Rscore.

\subsection{Modified Any Fit Evaluation}
\label{sub:exp_modified_any_fit}

Each algorithm was evaluated over a stream of measurements, persisting the
Rscore and the number of bins used for each iteration. The algorithms are
compared with respect to the same stream.

\begin{figure}[htb!] 
\centering
\includegraphics[width=0.9\columnwidth]{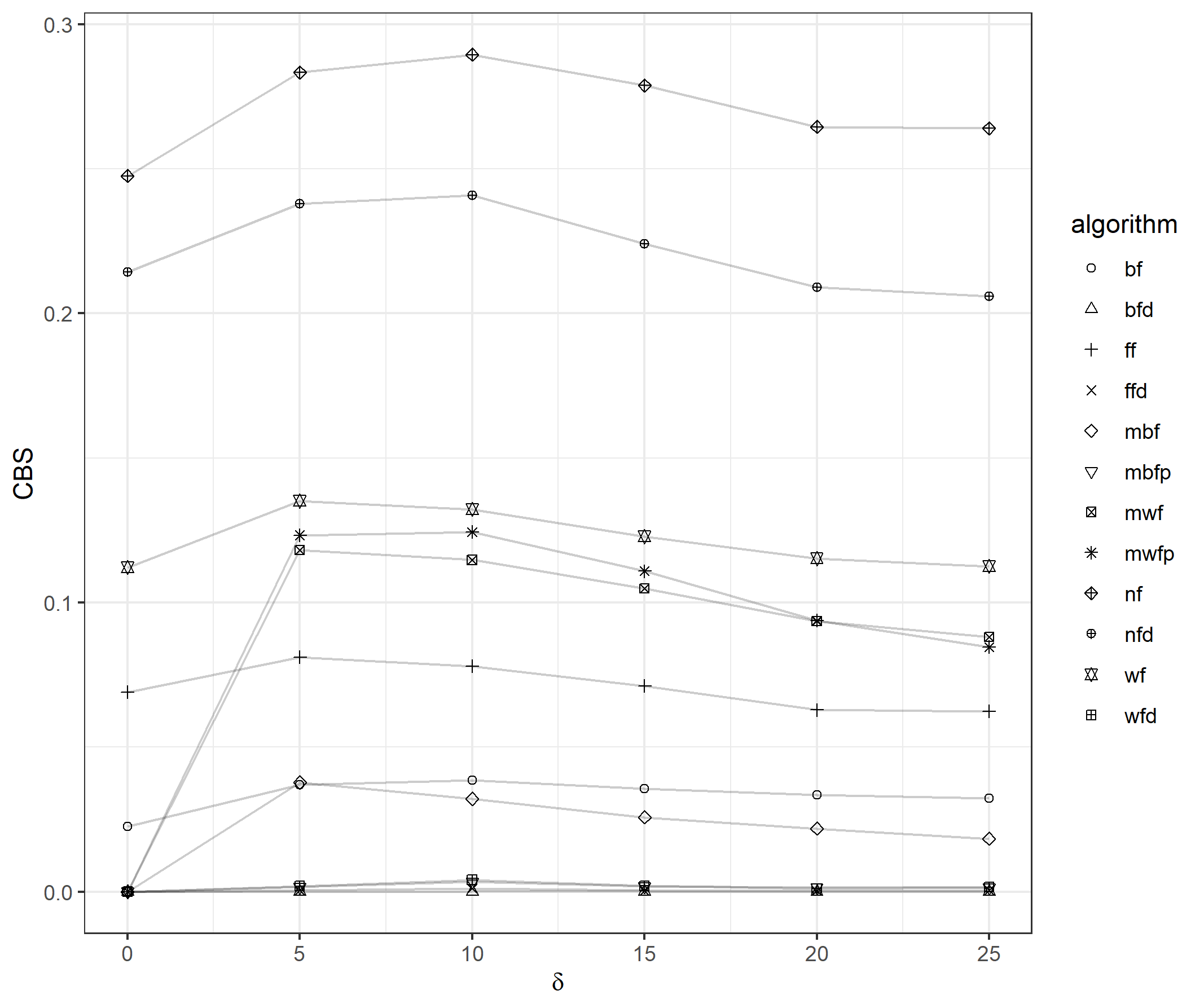} 
\caption{
    Cardinal Bin Score (CBS) for all implemented algorithms.
} 
\label{fig:relative_nconsumers} 
\end{figure}

When evaluating the Cardinal Bin Score, the worst performing algorithm, as shown
in Fig. \ref{fig:relative_nconsumers}, is the next fit followed by its
decreasing version. The remaining any fit decreasing algorithms, are the ones
that perform the best, with the best fit decreasing consistently presenting the
best results. 

\begin{figure}[htb!] 
\centering
\includegraphics[width=0.9\columnwidth]{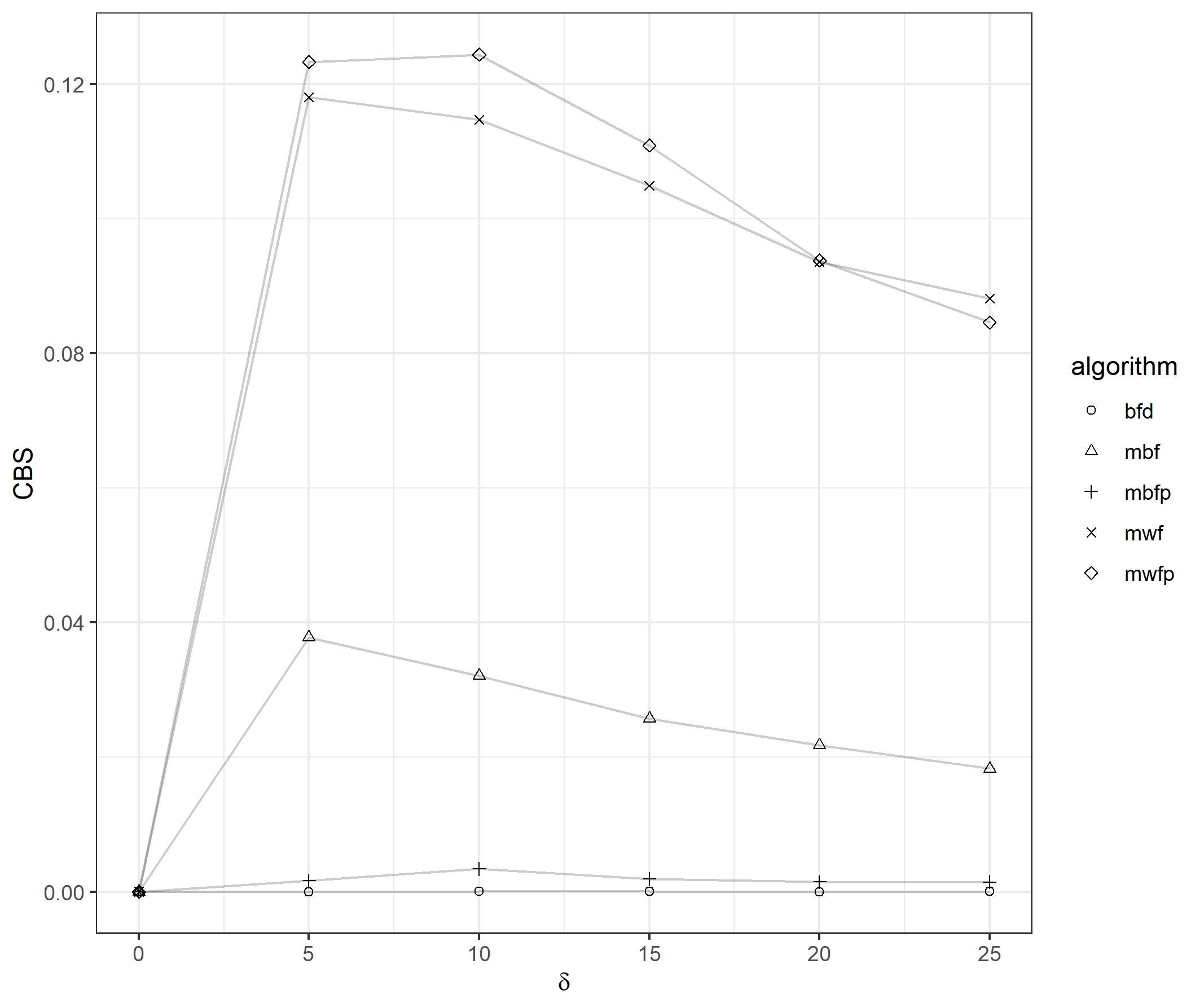} 
\caption{
    Cardinal Bin Score (CBS) filtered to present the modified and the
    BFD algorithms.
} 
\label{fig:relative_nconsumers_modified} 
\end{figure}

As for the modified versions of these algorithms, due to its sorting strategy,
MBFP shows the best results. It is also worth noting that for smaller
variabilities, the modified algorithms behave similarly to the online versions
of their any fit strategy with respect to the CBS, since the partitions aren't
necessarily assigned from biggest to smallest. On the other hand, the higher the
delta, the bigger the variability, which also leads to more rebalancing, having
the modified algorithms behave more like the decreasing versions of their fit
strategy.

\begin{figure}[htb!] 
    \centering
    \includegraphics[width=0.9\columnwidth]{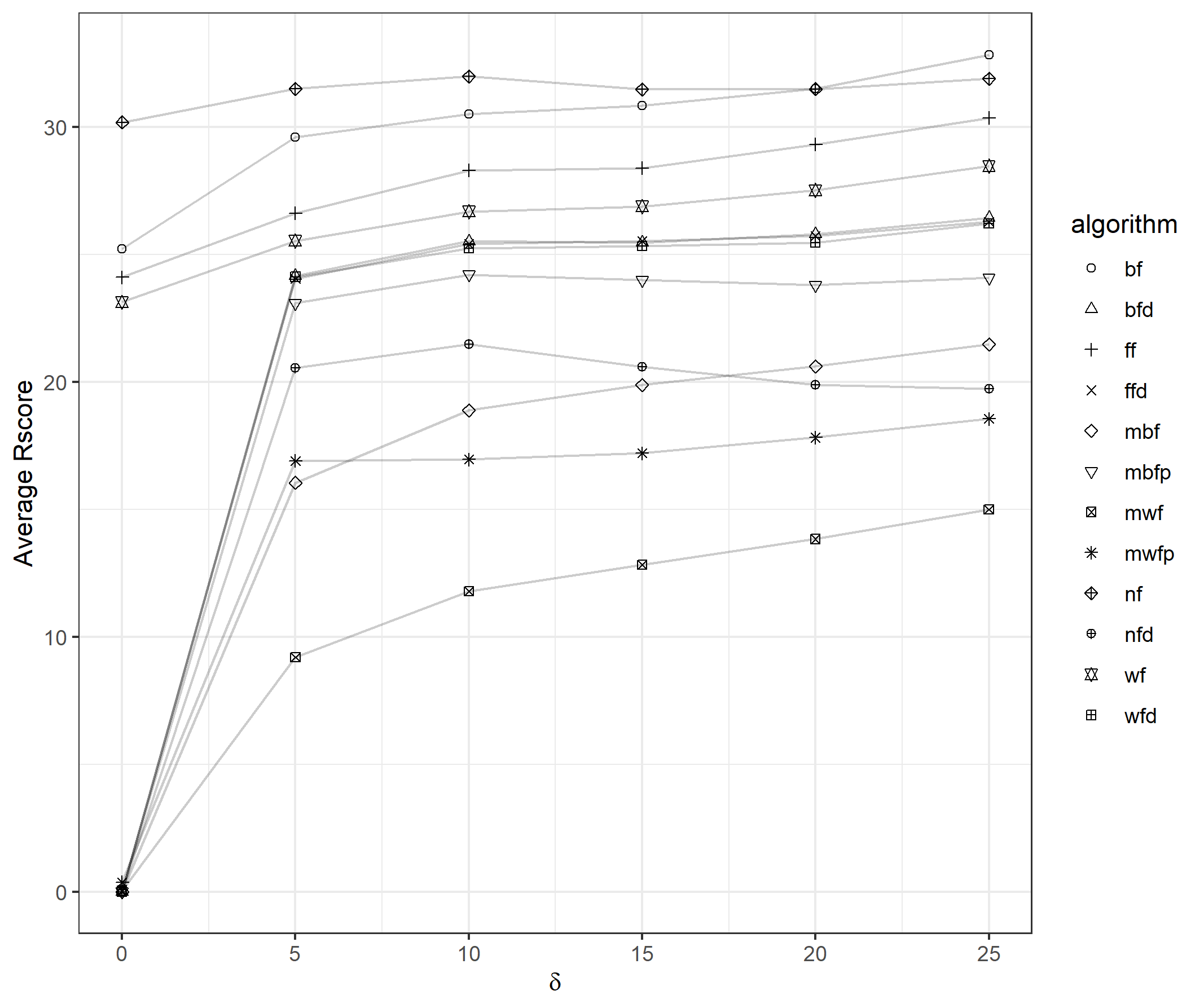}
    \caption{
        Impact on Rscore for different Deltas (random initial partition speed).
    } 
    \label{fig:rscore} 
\end{figure}

With regards to the Rscore, a common trend can be verified for all algorithms,
where an algorithm's Average Rscore increases with the value of $\delta$, as can
be seen in Fig. \ref{fig:rscore}. This should be expected as the bigger the
$\delta$, the more variability there is between the items' sizes in two
consecutive measurements of a stream. 

With respect to the average Rscore, the algorithms that perform the best are the
Modified Any Fit algorithms and the NFD. The reason why the NFD appears within
the five best algorithms, is due to the increased amount of bins required to
pack the items into its bins. This is related to the modification made to the
existing approximation algorithms, presented in Sec. \ref{sub:approximation_algorithms_adaptation}. This modification
leads to less rebalancing, as the partition might not have to be assigned a
different bin than it is currently assigned, if new bin is created for it.

For a similar reason, the modified algorithms that perform the best with regards
to the Rscore, are also the ones that perform worst (compared to the remaining
modified algorithms) when evaluating the CBS. As such, on account of the added
rebalance concern within the modified algorithms, these present an improvement
when it comes to the rebalance cost compared to the existing approximation
algorithms.

\begin{figure}[!htb] 
    \centering
		\subfloat[
        Pareto front for $\delta=5$.
        \label{fig:pareto_front_a}
    ]{
		\includegraphics[width=0.45\linewidth]{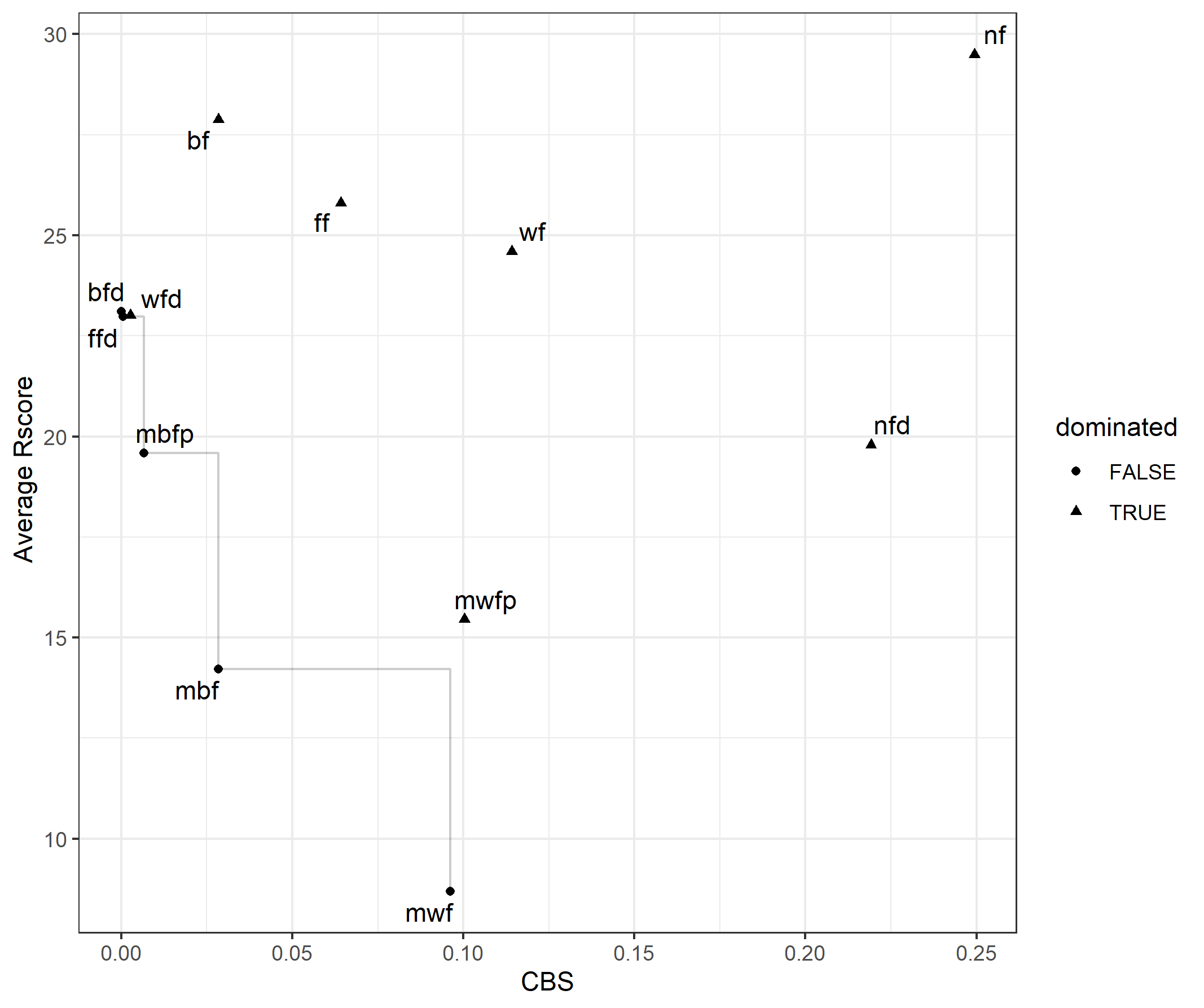}
    }
    \hfill
		\subfloat[
        Pareto front for $\delta=25$.
        \label{fig:pareto_front_b}
    ]{
        \includegraphics[width=0.45\linewidth]{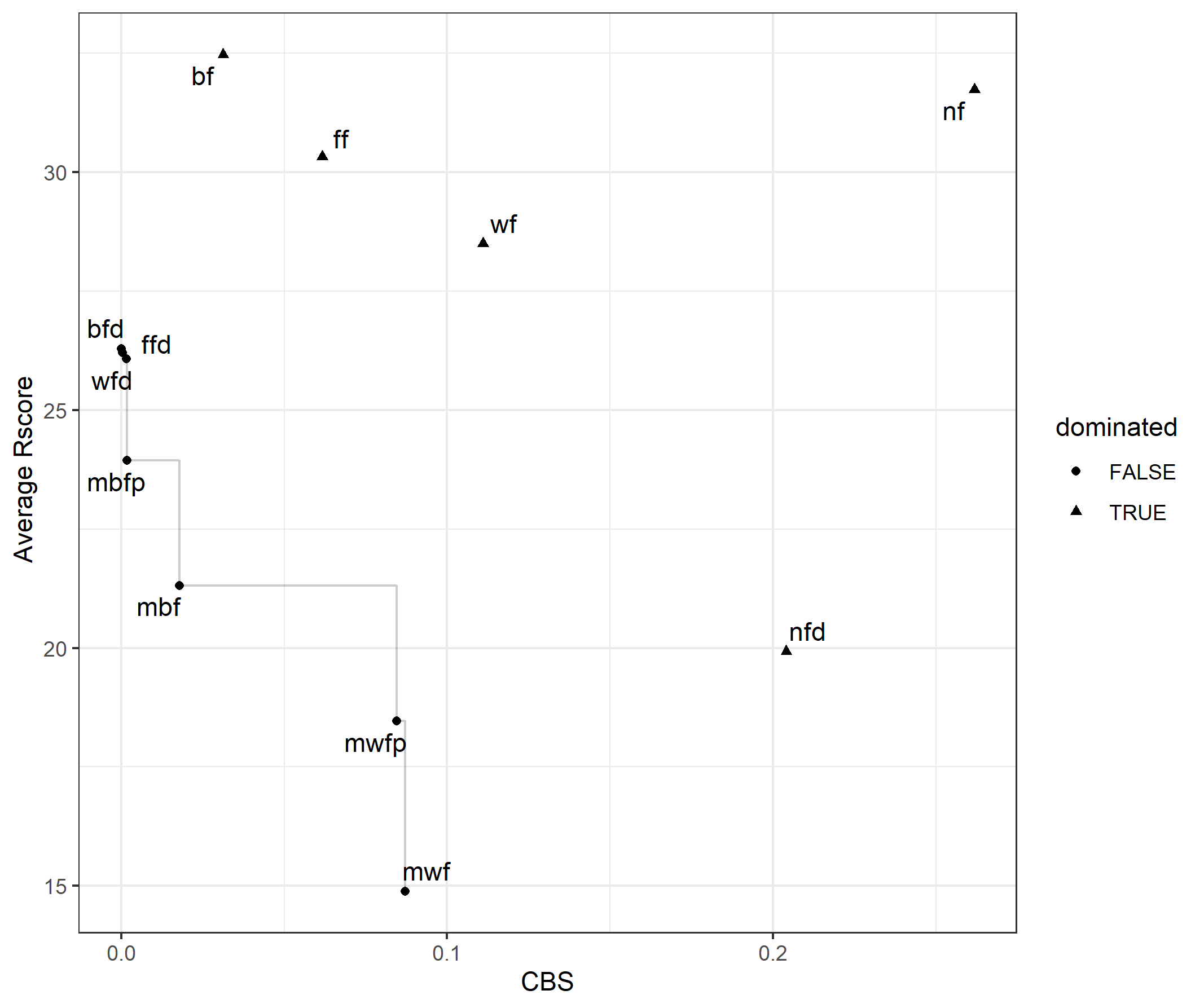}
    }
 		\caption{
        Pareto front for different deltas comparing the Cardinal Bin Score and
        the Average Rscore.
    }
\label{fig:pareto_front} 
\end{figure}

We find the pareto front, so as to determine the set of solutions that are most
efficient, provided there are trade-offs within a multi-objective optimization
problem. Excluding MWFP, the modified algorithms are consistently a part of the
pareto front, as shown in Fig. \ref{fig:pareto_front}, which implies these are
a competitive alternative when solving this variation of the BP problem.

\subsection{Kafka Consumer Group Autoscaler Evaluation}

The consumer group autoscaler, was developed as a proof of concept as to how
this variation of the Bin Packing problem could be applied in a message broker
environment. As such, we now present the results when running this system in
Maersk's (HUUB) production message broker environment, with the compute
resources summarized by Table \ref{tab:experimentation_resources}.

\begin{table}[H]
\caption{Infrastructure used for experimentation}
\label{tab:experimentation_resources}
\centering
\begin{tabular}{|p{0.18\columnwidth}|p{0.14\columnwidth}|p{0.5\columnwidth}|} 
    \hline
    \textbf{Component} 
        & \textbf{Cloud Platform} 
            & \textbf{Resource Classification}\\
    \hline
    
    Kafka Cluster (3 Nodes) 
        & AWS 
            & r5.xlarge \\
    \hline
    Kubernetes 
        & GCP
            & GKE autopilot \\
    \hline
    Monitor 
        & GCP 
            & Kubernetes Deployment (GKE) \\
    \hline
    Controller 
        & GCP  
            & Kubernetes Deployment (GKE) \\
    \hline
    Consumer Group
        & GCP 
            & As many Deployments as consumers in the group (GKE) \\
    \hline
\end{tabular}
\end{table}

\begin{figure}[htb!] \centering
\includegraphics[width=0.8\columnwidth]{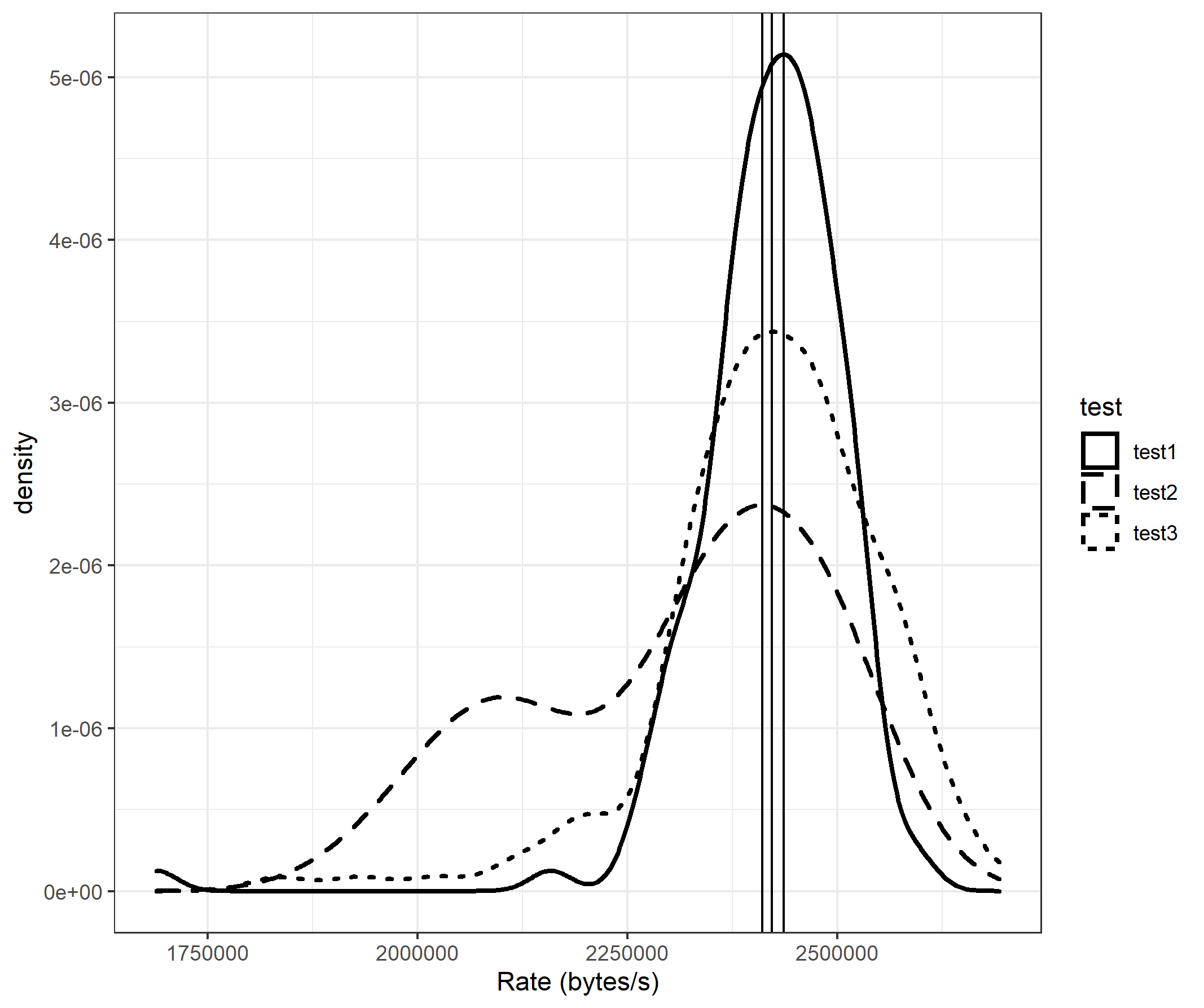}
\caption{
    Density Plot for the consumer's measured throughput in the three testing
    conditions.
} 
\label{fig:consumer_capacity} 
\end{figure}

One of the first assumptions, was that this problem was a variation of the SBSBP
problem, and therefore, the consumer should present a constant maximum capacity
when challenged to work at peak performance. To obtain this value, the consumer
was tested in three very disparate conditions, specified in Table \ref{tab:consumer_testing_conditions}, all
requiring the consumer to be reading the data at its maximum consumption rate.
Between each test, we varied: the average amount of bytes present in each
partition; the number of destination tables in the data lake; and the number of
partitions assigned to the consumer instance. Fig. \ref{fig:consumer_capacity} presents the results
of having the consumer run in these conditions, and clearly shows a common mode
for the maximum consumption rate, around $2.3 Mbytes/s$.

\begin{table*}[htb!] 
\centering 
\caption{
    Testing conditions to obtain consumer maximum throughput measure.
} 
\label{tab:consumer_testing_conditions}
    \begin{tabular}{ |c|r|r|r|r| } 
        \hline
        \textbf{Test ID} & \textbf{Total Bytes} & \textbf{Average Bytes} &
            \textbf{Number of Partitions} & \textbf{Number of Tables} \\ 
        \hline 
        Test 1 
            & $648\ Mbytes$ 
                & $20\ Mbytes$ 
                    & $32$ 
                        & $1$ \\ 
        Test 2 
            & $100\ Mbytes$ 
                & $0.86\ Mbytes$ 
                    & $116$ 
                        & $5$ \\ 
        Test 3 
            & $678\ Mbytes$ 
                & $4\ Mbytes$ 
                    & $144$ 
                        & $5$ \\ 
        \hline
    \end{tabular} 
\end{table*}

\section{Conclusions}
\label{sec:conclusions}

The work introduces a deterministic approach to
minimize the number of consumers required working in parallel, where message brokers are concerned. This minimizes operational costs. The approach assigns partitions to consumes guaranteeing that the rate of production into a set of partitions is not higher than the rate of consumption.

This study solves the consumer group autoscaling problem deterministically by analogizing the issue to a Bin Packing problem. The analogy corresponds item sizes to write speeds and bins to consumers. This paper also considers that write speeds can change over time, and assignments need to adapt to these changes. Reassignments pose a cost since they limit the consumers' capacity to read data. Terefore, we propose the Rscore to account for rebalancing costs.

To the best of our knowledge, there is no algorithm that accounts for the dynamic aspect of our BP problem. Hence, we propose four new BP-based heuristics that consider the re-scheduling costs (Rscore). Three of them are proven to be competitive solutions when solving the
multi-objective optimization problem. Thus, the assignments they lead to have some robustness to future write speed changes.


Lastly, this paper challenges some of Kafka's core functionalities and provides tested alternatives which can be used as a foundation so as to improve the way in which a consumer’s load is modeled, and the manner in which the load (partitions) is distributed between the elements of a consumer group. 
Ultimately, our approach solves a real-world problem regarding the sensor reads of a logistics company at lower operational costs than their previous non-functioning solution.

\bibliographystyle{IEEEtran}
\bibliography{refs}

\newpage


\end{document}